\documentclass[11pt]{article}
\def\iindex{ }
\def\iindex{\index }
\newcommand{\comm}[1]{}
\usepackage{amssymb}%,psfig}

\usepackage[numbers]{natbib}
\def\citet{\cite}
\newtheorem{theorem}{Theorem}[section]
\newtheorem{lemma}{Lemma}[section]

\newtheorem{proposition}{Proposition}[section]
\newtheorem{corollary}{Corollary}[section]

\newtheorem{definition}{Definition}[section]
\newtheorem{remark}{Remark}[section]

\newcounter{thanksnum}
\def\thanksnumber#1
{\setcounter{thanksnum}{\value{footnote}}\setcounter{footnote}{#1}%
                     \addtocounter{footnote}{-1}\footnotemark
                     \setcounter{footnote}{\value{thanksnum}}}
\def\newtheoremz#1{\@ifnextchar[{\@othmz{#1}}{\@nthmz{#1}}}

\def\@nthmz#1#2{%
\@ifnextchar[{\@xnthmz{#1}{#2}}{\@ynthmz{#1}{#2}}}

\def\@xnthmz#1#2[#3]{\expandafter\@ifdefinable\csname #1\endcsname
{\@definecounter{#1}\@addtoreset{#1}{#3}%
\expandafter\xdef\csname the#1\endcsname{\expandafter\noexpand
  \csname the#3\endcsname \@thmcountersepz \@thmcounterz{#1}}%
\global\@namedef{#1}{\@thmz{#1}{#2}}\global\@namedef{end#1}{\@endtheoremz}}}

\def\@ynthmz#1#2{\expandafter\@ifdefinable\csname #1\endcsname
{\@definecounter{#1}%
\expandafter\xdef\csname the#1\endcsname{\@thmcounterz{#1}}%
\global\@namedef{#1}{\@thm{#1}{#2}}\global\@namedef{end#1}{\@endtheoremz}}}

\def\@othmz#1[#2]#3{\expandafter\@ifdefinable\csname #1\endcsname
  {\global\@namedef{the#1}{\@nameuse{the#2}}%
\global\@namedef{#1}{\@thmz{#2}{#3}}%
\global\@namedef{end#1}{\@endtheoremz}}}

\def\@thmz#1#2{\refstepcounter
    {#1}\@ifnextchar[{\@ythmz{#1}{#2}}{\@xthmz{#1}{#2}}}

\def\@xthmz#1#2{\@begintheoremz{#2}{\csname the#1\endcsname}\ignorespaces}
\def\@ythmz#1#2[#3]{\@opargbegintheoremz{#2}{\csname
       the#1\endcsname}{#3}\ignorespaces}

\def\@thmcounterz#1{\noexpand\arabic{#1}}
\def\@thmcountersepz{.}
\def\@begintheoremz#1#2{ \trivlist \item[\hskip \labelsep{\bf #1\ #2}]}
\def\@opargbegintheoremz#1#2#3{ \trivlist
      \item[\hskip \labelsep{\bf #1\ #2\ (#3)}]}
\def\@endtheoremz{\endtrivlist}

\def\e{\varepsilon}

\def\defi{\stackrel{{\scriptscriptstyle \Delta}}{=}}

\def\OO{{\scriptscriptstyle O}}

\def\o{\omega}

\def\F{{\cal F}}
\def\w{\widehat}
\def\Ind{{\mathbb{I}}}

\def\sign{{\rm  sign\,}}

\def\const{{\rm const\,}}

\def\E{{\bf E}}
\def\P{{\bf P}}

\def\Z{{\cal Z}}

\def\b{\beta}
\def\s{\delta}
\def\g{\gamma}
\def\C{{\bf C}}

\def\ww{\widetilde}

\def\t{\theta}
\def\oo{\bar}
\def\s{\sigma}
\newcommand{\be}{\begin{equation}}
\newcommand{\ee}{\end{equation}}
\newcommand{\bd}{\begin{displaymath}}
\newcommand{\ed}{\end{displaymath}}
\newcommand{\ba}{\begin{array}{ll}}
\newcommand{\ea}{\end{array}}
\newcommand{\baa}{\begin{eqnarray}}
\newcommand{\eaa}{\end{eqnarray}}
\newcommand{\baaa}{\begin{eqnarray*}}
\newcommand{\eaaa}{\end{eqnarray*}}   \font\sm=cmr10
\def\OO{{\scriptscriptstyle \O}}
\def\k{\kappa}

\def\ww{\tilde}

\def\O{\Omega}

\def\ew{\left(e^{i\o}\right)}
\def\m0{m_{\scriptscriptstyle 0}}

\def\BL{\b_{\scriptscriptstyle min}}
\def\BL{{\scriptscriptstyle BL}}
\def\BL{{\scriptscriptstyle \O}}
\def\HH{{\bf H}}
\def\T{{\mathbb{T}}}
\def\ZZ{{\mathbb{Z}}}

\def\TT{{\cal \ZZ^-}}

\def\TT{{\cal T}}

\def\XN{\ell_2^{\scriptscriptstyle \O}}
\def\XNL{\ell_2^\BL(-\infty,0)}
\def\XL{\ell_2(-\infty,0)}
\def\BN{{\cal B}}%oldU_{\O ,N}
\def\BNN{{\cal B}}
\def\BN{{\mathbb{B}^\BL}}%oldU_{\O ,N}
\def\BNN{{\mathbb{B}^\BL}}
\title{ %On statistical undetectability of  the discrete time market incompleteness
On statistical indistinguishability of complete and incomplete  discrete time market models
}
\author{
Nikolai Dokuchaev\\
 {\sm Department of Mathematics \& Statistics, Curtin
University, 6845 Western Australia}\index{\\ {\sm  GPO Box U1987, Perth, 6845 Western Australia} }}
\begin{document}
\def\break{}%
\def\brea{}
\def\breakk{}
\maketitle
\begin{abstract} We investigate the possibility of statistical evaluation of the market completeness
 for discrete time
stock market models.  It is  known that the market completeness is
not a robust property: small random deviations of the coefficients
convert a complete market model into a incomplete one. The paper
shows that market incompleteness is also non-robust.
We show that, for any incomplete market  from a wide
class of discrete time models, there exists a complete market model with
arbitrarily close  stock prices. This means that incomplete markets are indistinguishable from the complete markets
 in the terms of the market statistics.
%C61 - Optimization Techniques; Programming Models; Dynamic Analysis
%G11% - Portfolio Choice; Investment Decisions
\index{{\bf MSC2010  classification}:
91G70,   %   Statistical methods, econometrics
91G20,   %  Derivative securities
91B84,   %   Economic time series analysis
91B26,   %  Market models (auctions, bargaining, bidding, selling, etc.)
62P05  %    Applications of statistics to actuarial sciences and financial mathematics
}\end{abstract}
 \section{Introduction}  The paper studies discrete time  stock market models and
 their completeness or incompleteness.  \iindex{ These concepts are
crucial for the modern mathematical finance.} For the so-called complete market, any claim can be replicated.
\iindex{ and
where there is a unique martingale (risk-neutral)  measure
equivalent to the historical measure. For complete market models,
the price of a derivative is defined via the expectation of the
payoff by this unique martingale measure.
 The classical
Black-Scholes continuous time market model with a non-random volatility is complete.}
The classical discrete time
Cox-Ross-Rubinstein  model of a single-stock  financial market is  complete; this is a binomial model.
For incomplete
market models, the option replication is not always possible. Unfortunately,
the market completeness is not a robust property:
small random deviations can ruin the completeness and convert a
complete model into a incomplete one.
\par
In the present paper, we \iindex{study  the robustness of market incompleteness.  We follow
the general approach to detecting the non-robustness of certain market properties
introduced by Guasoniy and  R\'asonyi in \cite{GR}, where non-robustness of arbitrage
opportunities was established.   We} show that the market incompleteness is also non-robust.   It appears that, for any incomplete market model from a wide class of
models, there exists a complete market model with an arbitrarily
close stock prices,  in a setting where the admissible
portfolio strategies  can use historical observations collected
 before the launching time of the replicating
strategy  (Theorem
\ref{ThM}). In other word, the incomplete
markets are indistinguishable from the complete markets in the terms
of the market statistics (Corollary \ref{corr1}). Arbitrarily small rounding errors
and time discretization errors may lead to different market models with
respect to the completeness and incompleteness. This contradicts to
 a common perception that the
incompleteness can be spotted from
the statistics.
\par
Theorem \ref{ThM} establishes some limits for analysis of market
structures based solely on econometrics and provides one more illustration of importance of the agents' beliefs in interpretations of
econometrical data, in the framework of the concept from
\cite{Madan}-\cite{ME}.  Another curious consequence is that the option prices are not robust with
  respect to small deviations of the past  stock prices, since pricing formulas for complete and incomplete
  models are different (in fact, prices are not uniquely defined for the incomplete market).
\par
Some non-robustness  of certain market properties (more precisely, arbitrage
opportunities) was considered in  \cite{GR}.  We
study  a different market property: the incompleteness  caused by
non-hedgeable randomness of parameters. The arbitrage possibility
or completeness are some extreme and rare features.  The arbitrage
possibility  is usually caused by abnormally vanishing volatility or
fast growing appreciation rate; the completeness is caused by the
predictability and the absence of the noise for the volatility. On
the other hand, the incompleteness is rather a typical feature.
It is easier to believe that a noise contamination of a model
removes some rare property. Hence the result of the present paper is  more
counterintuitive.
\par
Related results were  obtained in \cite{D12comp,D14} and presented
by the author on The Quantitative Methods in Finance conference in Sydney in 2013. In \cite{D12comp},
diffusion continuous time models were considered;   in \cite{D14}, discrete time  high frequency  binomial models and their were considered. The result of the present paper
 was obtained by a different approach.
\section{The result}
\subsection{The market model}
Assume that we are given a  probability space with  a complete $\s$-algebra of events $\F$ and  a
probability measure $\P$. Let $\ZZ$  be the set of all integers, and let $\ZZ^-=\{0,-1,-2,-3,...\}$.
\par

\index{Assume that we are given a  probability space
$(\Omega,\F,\P)$, where $\Omega$ is a set of elementary
events, $\F$ is a complete $\s$-algebra of events  and $\P$ is a
probability measure.}

Consider discrete time   model of a securities market
consisting of a risky stock with the price $S(t)>0$ and risk free bond or bank account with the price
$B(t)$, for integers $t$.  The process $B(t)$ is assumed to be non-random and such that $B(t)>0$ a.s.
For simplicity, we assume that $B(t+1)/B(t)=\rho$ for some $\rho\ge 1$. Let $\ww S(t)=B(t)^{-1}S(t)$ be the discounted price process.
 In this setting,  the process $B(t)$ is assumed to be non-random
or risk-free and is used as a num\'eraire.

Let $\{\F_t\}$ be the filtration generated by the flow of observable
data, i.e., by the process $S(t)$.

Let  $\xi(t)=(\ww S(t)/\ww S(t-1)-1)$.
Clearly,  $\ww S(t)=\ww S(t-1)(1+\xi(t))$.

We assume that $\xi(t)\in (-1,1)$. It can be noted that the presence of the upper boundary for $\xi(t)$ is actually
 restrictive since it excludes some important models; however,
our proof for the results given below depends on this assumptions.

We assume that there exists a probability measure $\P_*$ being equivalent to $\P$ such that the process $\ww S(t)$ is a martingale with respect to
$\{\F_t\}$. Let $\E_*$ be the corresponding expectation.

Let $s,\t\in\ZZ$ be given, $s< \t$.
Let $X(t)$ be the wealth at time $t$ and such  that
\begin{equation}
\label{3.3} X(t)=\b(t)B(t)+\g(t)S(t),\quad t=s,s+1,...,\t,
\end{equation}
where $\b(t)$ is the quantity of the bond portfolio and where
$\g(t)$ is the vector describing the quantities of the shares of the stock
portfolio. The pair $(\b(t), \g(t))$ describes the state of the
bond-stocks securities portfolio at time $t$. We call the
sequences of these pairs portfolio strategies.
\par
Some constraints will be imposed on current operations in the
market.
\par
%\begin{definition}\label{defAS}
A portfolio strategy  $\{(\b(t),\g(t))\}_{t=s}^{\t}$ is said to be  admissible and self-financing  if
the following conditions are satisfied.
\begin{itemize}
\item[(i)] There exists a $\P$-equivalent martingale measure $\P_*$ such that $\E_*\g(t)^2<+\infty$ and  $\E_*\b(t)^2<+\infty$ for $t=s,...,\t$.
\item[(ii)]
The process
$(\b(t),\g(t))$ is adapted to the filtration $\{\F_t\}$.
\item[(iii)] For $t=s,...,\t-1$, \baaa
 X(t+1)-X(t)=\b(t)\left(B(t+1)-B(t)\right)
+\g(t) \left(S(t+1)-S(t)\right).
\label{3.4} \eaaa
\end{itemize}
%\end{definition}
We do not impose additional conditions on
strategies such as transaction costs, bid-ask gap, restrictions on
short selling; furthermore, we assume that shares are divisible
arbitrarily, and that the current prices are available at the time
of transactions without delay. 
\par

\index{The process $\ww X(t)\defi B(t)^{-1}X(t)$ is called the discounted
wealth. \begin{proposition}
\label{XwwX} Let $\{X(t)\}$ be a sequence, and let the sequence
$\{(\b(t),\g(t))\}$ be an admissible portfolio strategy, where
$\b(t)=(X(t)-\g(t)S(t))B(t)^{-1}$. Then  the process $\ww X(t)$
evolves as
  $$ \ww X(t+1)-\ww X(t)=\g(t)(\ww S(t)-\ww S(t)).
$$
\end{proposition}
\par
{\it Proof of Proposition \ref{XwwX}} is standard (see, e.g.,
\citet{P}). \index{Pliska (1997)).}
\par
It follows from Proposition \ref{XwwX} that the sequence $\{\g(t)\}$
alone suffices to specify admissible portfolio strategy
$\{(\b(t),\g(t))\}$.}

\begin{definition}\label{defC} Let $s,q\in\ZZ$ be such that $s<\t$.  A market model is said to be
complete for the time interval $\{s,s+1,...,q\}$ if, for any
$\F_\t$-measurable random claim $\psi$,
 such that  $\E_*\psi^{2}<+\infty$, there exists  $\F_s$-measurable initial wealth $X(s)$ and an admissible self-financing strategy defined at the times
sequence $\{s,s+1,...,q\}$ such that $X(\t)=B(\t)B(s)^{-1}\psi$ a.s. (i.e, $B(\t)B(s)^{-1}\psi$ is replicable with this
strategy and this initial wealth).
\end{definition}

Under the assumptions of Definition \ref{defC},  $X(s)=\E_*\psi$, and this is the fair price at time $s$
of an option with the payoff $B(\t)B(s)^{-1}\psi$ at the expiration time $q$. This price is   uniquely
defined, as well as the martingale measure.

The classical
Cox-Ross-Rubinstein discrete time model of a single-stock  financial market is
covered by this definition  with $s=0$ and trivial $\s$-algebra $\F_0$. For this model,
$\xi(t)$ takes only two values, $-d_1$ and $d_2$, such that $d_k\in (0,1)$, $k=1,2$; see, e.g., \cite{Dbook}, Chapter 3.
\index{Moreover, a market model with  $d_k(t)$, $k=1,2$ being measurable with respect
to  $\F_{t-1}$, where $\{\F_t\}$ is the filtration generated by $S$.}
A trivial generalization  of  the classical
Cox-Ross-Rubinstein  model gives the follwing proposition.
\begin{proposition}\label{propCRR}  A market model  is complete in the sense of Definition \ref{defC}
 if
$\xi(t)$ takes only two random values, $-d_1$ and $d_2$, such that $d_k$ are $\F_s$-measurable and
$d_k\in (0,1)$ a.s., $k=1,2$.
\end{proposition}

The pricing of derivatives is usually more difficult for the so-called incomplete market models where
a martingale measure is not unique. Some important examples of
market incompleteness arise for a modification of the model described above where $d_k(t)$, $k=1,2$ are not measurable with respect
to  $\F_{t-1}$, i.e., binomial models  with dynamically
adjusted sizes (i.e., random sizes) of the binary increments; see,
e.g., \cite{AD}. \index{  Akyildirim {\em et al}  (2012),} These binomial  models are
incomplete.

Let $\TT$ be a given subset of $\ZZ^-$.

Starting from now, we will consider $t=0$ as the current time; we will assume   that
the observations of the prices are available for $t\in\TT$. Inevitably, to consider pricing problems for the options expiring at a time  $T>0$,
we have to rely on a hypothesis that the properties of the market that we established using the historical observations will somehow be carried forward to the future times $t>0$. Therefore, we will be considering completeness based on observed prices for negative times.

\begin{theorem}\label{ThM}
 Let $\{S(t)\}_{t\in\TT}$ be the set of prices for the model described above. \begin{itemize}
\item[(i)]
Let $\TT=\{t: \ \t\le t\le 0\}$, for some $\t<0$. In this case, for any $\e>0$, there exists a
market model with the corresponding processes $\{\ww S_{\e}(t)\}$ and  $\{\xi_{\e}(t)\}$ that
 is complete on the time interval $\{s,...,q\}$ for any $s,q\in\TT$, $s< q$, and such that \baa
\sup_{t\in\TT}\left(\left|\vphantom{1^A}S_\e(t)-S(t)\right|+\left|\vphantom{1^A}\xi_\e(t)-\xi(t)\right|\right)<\e \quad \hbox{a.s.}\label{eps}\eaa
\item[(ii)] Let  $\TT=\ZZ^-$ and let
 there exists $M>0$ such that $\sum_{t\in\TT}(1+|t|)^{-2M}|\xi(t)|^2<+\infty$ a.s..
In this case, for any $\tau<0$ and $\e>0$, there exists a market model with the corresponding processes $\{\ww S_{\e}(t)\}$ and  $\{\xi_{\e}(t)\}$ that is complete on the interval $\{s,...,q\}$ for any $s,q\in\TT$, $s< q$, and that \baa
&&\sum_{t\in\TT}(1+|t|)^{-2M}\left(\vphantom{1^A}\xi_\e(t)-\xi(t)\right)^2<\e \quad \hbox{a.s.},\nonumber\\
&&\sup_{t:\, \tau\le t\le 0}|S_\e(t)/S_\e(\tau)-S(t)/S(\tau)|<\e \quad \hbox{a.s.}\label{epsM}\eaa
\end{itemize}
\end{theorem}

\begin{corollary}\label{corr1}  The incomplete markets are
indistinguishable from the complete markets in the terms of the
market statistics.
\end{corollary}

\section{Proof of Theorem \ref{ThM}}
For $r\in [1,+\infty]$ and $\t,\tau\in\ZZ$, $\t\le\tau$, we denote by $\ell_r(\t,\tau)$ the Banach
space of real valued sequences $\{x(t)\}_{t=\t}^\tau$
with the norm
$\|x\|_{\ell_r(\t,\tau)}\defi\left(\sum_{t=\t}^\tau|x(t)|^r\right)^{1/r}$ for $r<+\infty$
and $\|x\|_{\ell_\infty(\t,\tau)}\defi\sup_{t}|x(t)|$ for $r=+\infty$.
Similar notations will be used for $\t=-\infty$ and $\tau=+\infty$.
In addition, for a  $\oo\TT\subset\ZZ$, we will use a similar notation  $\ell_2(\oo\TT)$ the Banach space  of
 real valued sequences $\{x(t)\}_{t\in\TT}$
with the norm
$\|x\|_{\ell_2(\t,\tau)}=\left(\sum_{t=\t}^\tau|x(t)|^2\right)^{1/2}$.

Let $\ell_r\defi\ell_r(-\infty,+\infty)$.
\par
For  $x\in \ell_2$, we denote by $X=\Z x$ the
Z-transform  \baaa X(z)=\sum_{t=-\infty}^{\infty}x(t)z^{-t},\quad
z\in\C. \eaaa The inverse Z-transform  $x=\Z^{-1}X$ is
defined as \baaa x(t)=\frac{1}{2\pi}\int_{-\pi}^\pi
X\left(e^{i\o}\right) e^{i\o t}d\o, \quad t=0,\pm 1,\pm 2,....\eaaa

We assume that we are given $\O\in(0,\pi)$.
\par
For a $\O\in(0,\pi)$, let $\BNN$ be the set of all mappings $X:\T\to\C$ such
that $X\ew \in L_2(-\pi,\pi)$ and $X\ew =0$ for $|\o|>\O$. We will call the the corresponding processes $x=\Z^{-1}X$
{\em band-limited}. Let $\XN$ be the set of all band-limited processes from $\ell_2$.

Let $H_\OO(z)$ be the transfer function for an ideal low-pass filter such that $H_\OO\ew=\Ind_{[-\O,\O]}(\o)$, where
$\Ind$ is the indicator function. Let $h_\OO=\Z^{-1}H_\OO$.

For a subset $\oo\TT\subset\ZZ^-$, let $\ell_2^\OO(\oo\TT)$ be the subset of $\ell_2(\oo\TT)$ consisting of sequences
$\{\w x(t)\}_{t\in\TT}$ for all $\w x\in\XN$. We will use notation $\ell_2^\BL(-\infty,0)=\ell_2^\OO(\ZZ^-)$.

\begin{lemma}\label{propU}
\begin{itemize}
 \item[(i)]  For any $\tau\in\ZZ$ and any $\w x\in\ell_2^\BL(\ZZ^-_\tau)$, where
 $\ZZ^-_\tau\defi\{t:\quad t\le\tau\}$, there exists an unique $x'\in\ell_2$ such that $\w x(t)=x'(t)$ for $t\le \tau$.
\item[(ii)] For any $\O\in (0,\pi)$,  the set $\ell_2^\BL(-\infty,0)$ is a closed linear subspace of $\ell_2(-\infty,0)$.
\item[(iii)] For any  $x\in\XL$, there exists an unique   projection $\w x_\OO$   on $\XNL$. In addition, for $r=2$ and $r=+\infty$,
\baaa \|x-\w x_\OO\|_{\ell_r(-\infty,0)}\to 0\quad\hbox{as}\quad \O\to\pi-0. \eaaa
\item[(iv)] If $\TT$ is a finite set, then
 $\{x(t)\}_{t\in\TT}\in \ell_2^\BL(\TT)$ for any $x\in\ell_2$, and there exist more than one  $\w x_\OO\in \XN$ such that $x(t)=\w x_\OO(t)$ for $t\in\TT$.
\end{itemize}
\end{lemma}
\par
 {\em Proof of Lemma \ref{propU}}.
Let us prove statement (i).  It suffices to consider $\tau=0$ only and
prove that if $x(\cdot)\in\XN $ is such that $x(t)=0$ for
$t\le 0$, then $x(t)=0$ for $t>0$.  By Theorem 1 from \citet{D12a},
processes $x(\cdot)\in\XN $ are weakly predictable in the
following sense: for any $T>0$ , $\e>0$, and $\k\in
\ell_\infty(0,T)$, there exists  $\w \k(\cdot)\in
\ell_2(0,+\infty)\cap \ell_\infty(0,+\infty)$ such that
$
\|y-\w y\|_{\ell_2}\le \e,
$
where
 \baaa &&y(t)\defi
\sum_{m=t}^{t+T}\k(t-m)x(m),\qquad\breakk \w y(t)\defi
\sum^{t}_{m=-\infty}\w \k(t-m)x(m).\label{predict} \eaaa
We apply this to a process $x(\cdot)\in\XN $ such that
$x(t)=0$ for $t\in\ZZ^-$. Let us observe first that \baa \w
y(t)=0\quad \forall t<0. \label{zero} \eaa Let $T>0$ be given. Let
us show that $x(t)=0$ if $0\le t\le T$.  Let
$\{\k_i(\cdot)\}$ be a basis in $\ell_2(-T,0)$. Let
$y_i(t)\defi\sum_{m=t}^{t+T} \k_i(t-m)x(m)$. It follows from
(\ref{zero}) and from the weak predictability \citet{D12a,D12b} of $x$  that $y_i(t)=0$ if $t\le 0$.  It
follows that $x(t)=0$ if $t\le T$.
\par
Further, let us apply the proof given above to the process
$x_T(t)=x(t+T)$. Clearly, $x_T(\cdot)\in\XN $ and $x_1(t)=0$
for $t<0$. Similarly, we obtain that $x_T(t)=0$ for all $t\le T$,
i.e., $x(t)=0$ for all $t<2T$. Repeating this procedure $n$ times,
we obtain that $x(t)=0$ for all $t<nT$ for all $n\ge 1$. This
completes the proof of Lemma \ref{propU}(i). In particular, it follows that there exists
   $\w X\in\BN$ such that $\w x(t)=(\Z^{-1}\w X)(t)$ for $t\le 0$.

To prove statement (ii), it
suffices to  prove that $\XNL $ is a closed linear subspace of
$\ell_2(-\infty,0)$.
Consider the mapping $\zeta:\BN \to \XNL$ such that
$x(t)=(\zeta (X))(t)=(\Z^{-1} X)(t)$ for $t\in\ZZ^-$. This is a linear
continuous operator. By Lemma \ref{propU}(i), it is a bijection.
In this case,  there exists a unique projection $\w
x$ of $\{x(t)\}_{t\in\ZZ^-}$ on $\XNL$.
\par
 Since  the mapping $\zeta:\BN \to \XNL$ is continuous, it follows that
the inverse mapping $\zeta^{-1}: \XNL\to\BN$ is also
continuous; see, e.g., Corollary in Ch.II.5 \citet{Yosida}, p. 77. Since the
set $\BN$ is a closed linear subspace of $L_2(-\pi,\pi)$, it
follows that $\XNL$ is a closed linear subspace of $\ell_2(-\infty,s)$.
 This completes the proof of  statement (ii).

Let us prove statement (iii).  Let $X=\Z (x\Ind_{\ZZ^-})$ and $\ww X_\OO=H_\OO X$. Clearly,
\baaa
\|\w x_\OO- x\|_{\XL}&\le &\|\Ind_{\ZZ^-} h_\OO\circ (x\Ind_{\ZZ^-})- x\Ind_{\ZZ^-}\|_{\XL}
\\&\le& \const\|\ww X_\OO\ew- X\ew\|_{L_2(-\pi,\pi)}\to 0\quad \hbox{as}\quad
\O\to \pi.
\eaaa
This completes the proof of  statement (iii).

Let us prove statement (iv). \index{ can take from foresind?} Let us select arbitrarily
$q\in\ZZ^-\backslash \TT$. Let $\ww\TT=\TT\cup\{q\}$. Consider a finite system of equations
\baa
x(t)=\frac{1}{2\pi}\int_{-\O}^\O
\ww X\left(e^{i\o}\right) e^{i\o t}d\o, \quad t\in\ww\TT.
\label{wX}\eaa
Let us show that there exists $\ww X\ew\in L_2(-\O,\O)$ satisfying this system.
Consider  a set of linearly independent functions $\{\phi_m\}_{m\in\ww\TT}$ from   $L_2(-\O,\O)$
such that
\baaa\int_{-\O}^\O
\phi_m(\o)e^{i\o t}d\o=0, \quad t\in\ww\TT\backslash\{ m\}, \quad\qquad \int_{-\O}^\O
\phi_m(\o)e^{i\o m}d\o\neq 0.
\eaaa
In this case,  $\ww X\ew =\sum_{m\in\ww\TT}^0c_m\phi_m(\o)$ satisfy system (\ref{wX}) if  $c_m= \left(\int_{-\O}^\O
\phi_m(\o)e^{i\o m}d\o\right)^{-1}x(m)$.
  Let $X\ew=\ww X\ew$ for $\o\in[-\O,\O]$ and $X\ew=0$ for $\o\in[-\pi,\pi]\backslash[-\O,\O]$. The process $\w x_\OO=\Z^{-1}X$ is band-limited and has the desired values $x(t)$ for $t\in\TT$.
Clearly, these processes $\w x_\OO$ are different for different selections of  $x(q)$.
This completes the proof of statement (iv) and the proof of Lemma \ref{propU}.
\par
\begin{remark}
Lemma \ref{propU}(i) implies that the future $\{\w x(t)\}_{t>0}$ of a
band-limited   process
 is uniquely defined by its  past
$\{\w x(t)\}_{t\le 0}$.   This is a reformulation in the deterministic setting
of  the classical Szeg\"o-Kolmogorov Theorem established for stationary Gaussian processes
\citet{Sz,Sz1,K}.
Lemma \ref{propU}(iv) implies that if  $\TT$ is a finite set then any path  $\{x(t)\}_{t\in\TT}$ is a trace of a band-limited process.
 \end{remark}
 \par
 We now in the position to prove Theorem \ref{ThM}.
For the case of Theorem \ref{ThM}(i), we assume below that $M=0$.

Let $x(t)\defi(1+|t|)^{-M} |\xi(t)|$, and let $\w x_\OO(t)$  be the corresponding band-limited process described  in Lemma \ref{propU}(iii) if $\TT=\ZZ^+$ or any process  described  in Lemma \ref{propU}(iv) if $\TT$ is finite.
By Lemma \ref{propU}(iii),(iv), for any $\e_1>0$, there exists $\O=\O(\e_1)\in(0,\pi)$  such that
 \baa
\sup_{t\in\TT}\left|\w x_\OO(t) -x(t)\right|=
\sup_{  t\in\TT}\left|\w x_\OO(t) -(1+|t|)^{-M} |\xi(t)|\right|<\e_1 \quad \hbox{a.s.}\label{e1}\eaa

Let the  $\sign$ function be defined as $\sign(x)=1$ for $x\ge 0$ and  $\sign(x)=-1$ for $x<0$.

Consider a market model similar to the one described above and with the stock  prices
$S_\e(t)$ such that   \baaa
&&\ww S_\e(t)=\ww S_\e(t-1)(1+\xi_\e(t)), \quad  \ww S_\e(t)=B(t)^{-1}S_\e(t),\quad t\in\TT,\\
&&\ww S_\e(t)=\ww S_\e(t),\qquad t<\t,\quad \TT=\{\t,...,0\}\neq \ZZ^-,
\eaaa
where \baa
\xi_\e(t)\defi \zeta(t)a_\e(t),  \quad \zeta(t)\defi\sign(\xi(t)), \quad
a_\e(t)\defi(1+|t|)^M \w x_\OO(t).\label{se}
\eaa
Here
 $\ww S_\e(t)$ is the discounted price process. The process of bond prices $B(t)>0$ is such as described above, i.e., it is non-random and such that  $B(t+1)/B(t)=\rho$ for some $\rho\ge 1$.
  \par
  Clearly, for any $s,t\in\TT$, $s<t$,
  \baaa
  \ww S(t)=\ww S(s)\prod_{k=s}^{t-1}(1+\xi(k+1)),\quad \ww S_\e(t)=\ww S_\e(s)\prod_{k=s}^{t-1}(1+\xi_\e(k+1)),
  \eaaa
The process $\w x_\OO$ is band-limited, hence it is predicable in the sense of
Lemma \ref{propU}(i). It follows that  the process $a_\e(t)$ is also predictable in the sense of
Lemma \ref{propU}(i).  Clearly,   $|\xi_\e(t)|=a_\e(t)$,  and the process  $|\xi_\e(t)|$ is also predictable, i.e.,
$|\xi_\e(t)|$ is  $\F_{\tau}$-measurable for any $\tau<t\le 0$. one can select  $\O$ such that (\ref{eps2}) holds and that (\ref{eps2}) implies (\ref{eps}).
Hence the market model with the stock price $S_\e(t)$ and the bond price $B(t)$ is complete in the sense of Definition \ref{defC}

  Let $\e>0$ and $\tau<0$ be given; we assume that $\tau=\t$ under the assumptions of Theorem \ref{ThM}(i).  Clearly,
  there exist $\e_1=\e_1(\e,\tau)>0$  and $\O=\O(\e_1)$ such that (\ref{epsM}) and (\ref{e1}) hold
and
\baa &&\sup_{t: \, \tau\le t\le 0}\left|\vphantom{1^A}\xi_\e(t)- \xi(t)\,\right|\le \e,
\nonumber\\ &&\sup_{t: \, \tau\le t\le 0}\left|\prod_{k=s}^{t-1}(1+\xi_\e(k+1))-\prod_{k=s}^{t-1}(1+\xi(k+1))\right|\le \e.
\label{eps2}
\eaa
Then  (\ref{epsM})  follows. 
\index{ This completes the proof of Theorem \ref{ThM} for the case where $\TT=\Z^-$. The proof
the case where $\TT\neq\Z^-$ follows from the proof given above, since requirement  (\ref{eps}) is weaker
for $\TT\neq \ZZ^-$. For the general case where
$\TT\neq\Z^-$, it sufficient to consider the model such as described above
with $\xi_\e$ defined as above but  with $S_\e(t)=S(t)$? $\xi(t)=0$? for $t\in\ZZ^-\backslash\TT$. Nado $S_\e\sim S?$
\vspace{3cm}}
This completes the proof of Theorem \ref{ThM}.

\begin{remark} The predictability of band-limited processes
used in the proof of Theorem \ref{ThM} does not
require optimality of the projection $\w x$.
For example, one can use an ideal low-pass
filter applied to $x$ arbitrarily extended on $t>0$. Furthermore,
filters with the exponential energy decay  also transfers processes into predictable ones
 \cite{D12a}. Therefore, these filters with the exponential
energy can be used  in the proof of Theorem \ref{ThM} instead of the low-pass filters.
\end{remark}
\section{Discussion}

Theorem \ref{ThM}  leads to a counterintuitive conclusion that the incomplete
markets are indistinguishable from the complete markets by
econometric methods, i.e., in the terms of the market statistics.
Due to rounding errors, the statistical indistinguishability leading
to this conclusion cannot be fixed via the sample increasing since
the statistics for the incomplete market models can be arbitrarily
close to the statistics of  the alternative complete models.

It can be elaborated as the following.   Assume that we
collect the marked data (the sequence of the prices) for
$t\le 0$, with the purpose to test the following
hypotheses $\HH_0$ and $\HH_A$ about the stock price evolution:
\index{\begin{itemize}
\item[$\HH_0$:]   the market is  incomplete; and
\item[$\HH_A$:]   the market is  complete.
\end{itemize}}
\begin{itemize}
\item[$\HH_0$:] the values $\{|\xi(t)|\}_{t\le 0}$ do not represents a path of a predictable process  (i.e., the market is  incomplete); and
\item[$\HH_A$:] the values $\{|\xi(t)|\}_{t\le 0}$    represent a path of a predictable process, i.e.
$|\xi(t)|$ are  $\F_\tau$-measurable for any $\tau<t$ (i.e., the market is  complete).
\end{itemize}

In these hypotheses, we consider only the properties of the "past" market, leaving
aside the speculations about the future properties; this would require
additional hypotheses about connections between past observations and the future scenarios.

According to Theorem \ref{ThM},  it is impossible to reject
 hypothesis $\HH_A$ based solely on the market prices collected.
 Due to rounding errors, the statistical indistinguishability leading
to this conclusion cannot be fixed via the sample increasing since
the statistics for the incomplete market models can be arbitrarily
close to the statistics of  the alternative complete models.
This implies that
the commonly accepted  selection of a incomplete model is not actually based on the
market statistics.  However, this selection is justified since
it stays in the accordance
with general acceptance of the
immanent non-predictability of the real world. For instance, we
would rather accept a model with the possibility of the
unpredictable jumps for the volatility than a model where these
jumps can be predicted, even if the statistical data supports both
models equally.

Further,  it is  known that the market completeness is
not a robust   property: small deviations of the observed binomial prices
convert a complete market model into a incomplete one. Thanks to
Theorem \ref{ThM} and approximation scheme described above, we can claim  now  that market incompleteness is
also non-robust: small deviations can convert an incomplete model
into a complete one. More precisely, it implies that, for any
incomplete market from a wide class of models, there exists a
complete market model with arbitrarily close discrete sets of the observed prices.

We do not consider approximating models where the values $\xi_\e(t)$ are  predictable, since these models allow
 and are inconsistent with reasonable systems of market agents' beliefs. arbitrage.
 In the proofs, we used models where $|\xi_\e(t)|$ are predicable; these models are
 arbitrage free  and can be consistent with reasonable systems  of agents' beliefs.

Unfortunately, the predictability of
$|\xi_\e(t)|$ used in the proof of Theorem \ref{ThM} to
set  an alternative complete model cannot be applied to  option pricing under the "natural"
hypothesis $\HH_0$. The stock returns   $\xi(t)$ and $\xi_\e(t)$   are pathwise close under
these hypotheses $\HH_0$ and $\HH_A$  for $t\le 0$; however, their properties
are quite different with respect to the predicability, and the future paths of  $\xi(t)$ and $\xi_\e(t)$ will not be necessarily close.
Moreover, since the new and the old models
produce arbitrarily close sets of prices, an observer, due the rounding error, cannot tell
apart these models with certainty, i.e., she cannot tell which model
generates the observed data. Effectively, the process
$|\xi_\e(t)|_{t\le 0}$ in the new model is not observable at time
$t=0$ for an observer from the old model.

\par
It can be noted that we can replace the hypothesis $\HH_0$ by a
hypothesis assuming a particular incomplete market model such as a
Markov chain model, etc.

 \iindex{\subsection*{Acknowledgment} This work  was supported by ARC grant of Australia DP120100928 to the
author.}


\begin{thebibliography}{100}
\bibitem{A1}
A\"it-Sahalia, Y., and Mykland, P. (2004). Estimating diffusions
with discretely and possibly randomly spaced data: A general theory.
{\em Annals of Statistics} {\bf 32}, 2186-2222.
\bibitem{AD}
Akyildirim, E.,
Dolinsky, Y.    Soner, H.M. (2014). Approximating stochastic
volatility by recombinant trees. {\em
 Annals of Applied Probability} {\bf 24}, 2176--2205.
 \bibitem{ab}
Andersen, T. G. and Bollerslev, T. (1998). Answering the skeptics:
Yes, standard volatility models do provide accurate forecasts. {\em
International Economic Review} {\bf 39}, pp. 885--905.
\bibitem{ab1}
 Andersen, T.G., Bollerslev, T.,
 Diebold, F.X., and Ebens, H. (2001).  The distribution of realized stock return volatility. {\em  Journal of
Financial Economics} {\bf 61}, pp. 43–76.
\bibitem{AB2}
Andersen, T.G., Bollerslev, T., Diebold, F.X., and Labys, P. (2003).
Modeling and forecasting realized volatility. {\em Econometrica} {\bf 71},
pp. 579--625
\bibitem{B}
Barndorff-Nielsen, O.E., Graversen S.E. and  Shephard, N. (2003),
Power variation \& stochastic volatility: a review and some new
results. {\em  Journal of Applied Probability} 41A, 133--143.
\index{\bibitem{D2002}  Dokuchaev N.G. (2002). {\it Dynamic portfolio
strategies: quantitative methods and empirical rules for incomplete
information.}
 Kluwer Academic Publishers, Boston.}
\bibitem{Dbook}
Dokuchaev N.G. {\it Mathematical finance: core theory, problems, and
statistical  algorithms}. Routledge, London and New York, January 2007, 209p.

\bibitem{D2010a}
 Dokuchaev, N. (2010). Predictability on finite horizon
for processes with exponential decrease of energy on higher
frequencies.
 {\it Signal processing} {\bf 90}, iss. 2,  696--701.
\bibitem{D12a}
 Dokuchaev, N. (2012).
On sub-ideal causal smoothing filters.
 {\it Signal Processing} {\bf 92}, iss. 1, 219-223.
\bibitem{D12b}   Dokuchaev, N. (2012). Predictors for discrete time processes with
energy decay on higher frequencies. {\em IEEE
Transactions on Signal Processing} {\bf 60}, No. 11, 6027-6030.
\bibitem{D12c}
 Dokuchaev, N. (2012). On predictors for band-limited and
high-frequency time series. {\em Signal Processing} {\bf 92}, iss.
10, 2571-2575.
\bibitem{D12comp}
 Dokuchaev, N. (2012).  On statistical indistinguishability of the
complete and incomplete markets, preprint,  arXiv:1209.4695. %
%\x\xRN: http://ssrn.com/abstract=2149951.
\bibitem{D14}
 Dokuchaev, N. (2014). On strong causal binomial approximation for stochastic processes. {\em Discrete and Continuous Dynamical Systems --
Series B (DCDS-B)} {\bf 20}, No.6, 1549--1562.
\bibitem{eh}
Elliott, R.J., Hunter, W.C., and Jamieson, B.M. (1998). Drift and
volatility estimation in discrete time. {\em  Jour. of Economic
Dynamics \& Control} {\bf 22}, 209-218.
\bibitem{GR} Guasoni, P. and  R\'asonyi, M.
(2012). Fragility of arbitrage and bubbles in diffusion models.
Working paper, http://ssrn.com/abstract=1856223.
\bibitem{hw} Hull,
J., and White, A. (1987). The pricing of options on assets with
stochastic volatilities. {\em  Journal of Finance} 42, 381--400.

\bibitem[Kolmogorov  (1941)]{K}
 Kolmogorov, A.N. (1941).
 Interpolation and extrapolation of stationary stochastic series.
 {\em Izv. Akad. Nauk SSSR Ser. Mat.,} 5:1, 3--14.
\bibitem{Madan} Madan D.B. (1983).
Inconsistent Theories as Scientific Objectives. {\em Philosophy of
Science}, Vol. 50, No. 3, pp. 453--470.
\bibitem{ME}
Madan, D.B., and Eberlein, E. (2012). Dealing with complex realities
in financial modeling. {\em Current science} {\bf 103} (6),
647--649.
\bibitem{mm}
Malliavin, P., and Mancino, M.E. (2002). Fourier Series method for
measurement of multivariate volatilities. {\em Finance \&
Stochastics} 6, 49-62.
\bibitem[Pliska (1997)]{P} Pliska, S. R. (1997). {\it
Introduction to mathematical finance: discrete time models. }
Blackwell Publishers,  Oxford, UK, and Malden, Mass.
\bibitem[Szeg\"o (1920)]{Sz}
Szeg\"o, G. (1920). Beitr\"age zur Theorie der Toeplitzschen
Formen. {\em Math. Z.} 6, 167--202.
\bibitem[Szeg\"o (1921)]{Sz1}
 Szeg\"o, G. (1921). Beitr\"age zur Theorie der Toeplitzschen Formen, II. {\em Math. Z.}
9, 167-190.
\bibitem[Yosida (1965)]{Yosida} Yosida, K. (1965). {\em
Functional Analysis.}  Springer, Berlin Heilderberg New York.
\end{thebibliography}
\end{document}